\documentclass[preprint,showpacs,amsmath,amssymb]{revtex4}
\usepackage{graphicx}
\usepackage{dcolumn}
\usepackage{bm}
\newcommand{\be}{\begin{eqnarray}}
\newcommand{\en}{\end{eqnarray}}
\newcommand{\ben}{\begin{eqnarray*}}
\newcommand{\enn}{\end{eqnarray*}}
\newcommand{\pa}{\partial}

\newcommand{\f}{\frac}

\newcommand{\p}{\paragraph{}}
\newcommand{\bi}{\begin{itemize}}
\newcommand{\ei}{\end{itemize}}

\newcommand{\la}{\langle}
\newcommand{\ra}{\rangle}

\newcommand{\R}{\Rightarrow}
\renewcommand{\r}{\rho}
\renewcommand{\p}{\bot}
\renewcommand{\a}{\alpha}
\renewcommand{\b}{\delta}
\renewcommand{\omega}{q}
\newcommand{\mega}{Q}
\renewcommand{\eta}{\varepsilon_q}
\begin{document}
%%%%%%%%%%%%%%%%%%%%%%%%%%%%%%%%%%%%%%%%%%%%%%%%%%%%%%%%%%%%%%%%%%%%%%%%%%%%%%%%%%%%%%%%%%%%%%%%%%%%%%%%%%%%%%%%%%%%%%%%%%%%
\title{Two point third order correlation functions for quasi-geostrophic turbulence: Kolmogorov-Landau approach}
%%%%%%%%%%%%%%%%%%%%%%%%%%%%%%%%%%%%%%%%%%%%%%%%%%%%%%%%%%%%%%%%%%%%%%%%%%%%%%%%%%%%%%%%%%%%%%%%%%%%%%%%%%%%%%%%%%%%%%%%%%%%
\author{Sagar Chakraborty}
\email{sagar@bose.res.in}
\affiliation{S.N. Bose National Centre for Basic Sciences\\Saltlake, Kolkata 700098, India}
%%%%%%%%%%%%%%%%%%%%%%%%%%%%%%%%%%%%%%%%%%%%%%%%%%%%%%%%%%%%%%%%%%%%%%%%%%%%%%%%%%%%%%%%%%%%%%%%%%%%%%%%%%%%%%%%%%%%%%%%%%%%
\date{\today}
%%%%%%%%%%%%%%%%%%%%%%%%%%%%%%%%%%%%%%%%%%%%%%%%%%%%%%%%%%%%%%%%%%%%%%%%%%%%%%%%%%%%%%%%%%%%%%%%%%%%%%%%%%%%%%%%%%%%%%%%%%%%
\begin{abstract}
We use the more intuitive approach due to Kolmogorov (and subsequently, Landau in his text on fluid dynamics) to calculate some third-order structure functions for quasi-geostrophic turbulence for the forward cascade of pseudo-potential enstrophy and the inverse energy cascade in quasi-geostrophic turbulence.
\end{abstract}
%%%%%%%%%%%%%%%%%%%%%%%%%%%%%%%%%%%%%%%%%%%%%%%%%%%%%%%%%%%%%%%%%%%%%%%%%%%%%%%%%%%%%%%%%%%%%%%%%%%%%%%%%%%%%%%%%%%%%%%%%%%%
\pacs{47.27.-i, 92.60.hk, 92.10.Lq}
%%%%%%%%%%%%%%%%%%%%%%%%%%%%%%%%%%%%%%%%%%%%%%%%%%%%%%%%%%%%%%%%%%%%%%%%%%%%%%%%%%%%%%%%%%%%%%%%%%%%%%%%%%%%%%%%%%%%%%%%%%%%
\maketitle
%%%%%%%%%%%%%%%%%%%%%%%%%%%%%%%%%%%%%%%%%%%%%%%%%%%%%%%%%%%%%%%%%%%%%%%%%%%%%%%%%%%%%%%%%%%%%%%%%%%%%%%%%%%%%%%%%%%%%%%%%%%%
%
\section{Introduction}
Quasi-geostrophic (QG) turbulence is a rather more realistic class of turbulent flow than the isotropic homogeneous 3D turbulence.
It can be seen in the large scale flows on oceans and atmosphere; thus having profound geophysical and astrophysical significance.
In the incompressible isotropic homogeneous 3D turbulence, Kolmogorov's four-fifths law\cite{Kolmogorov} is a landmark in the theory of turbulence because it is a rare exact result.
In three spatial dimensions, this law says that the third order velocity correlation function behaves as:
\be
\left<\left[\left\{\vec{u}(\vec{x}+\vec{r})-\vec{u}(\vec{r})\right\}.\f{\vec{r}}{|\vec{r}|}\right]^3\right>=-\f{4}{5}\varepsilon r
\en
where $\varepsilon$ is the rate per unit mass at which energy is being transferred through the inertial range.
The inertial range is the intermediate spatial region postulated by Kolmogorov where the large scale disturbances (flow maintaining mechanisms) and the molecular scale viscous dissipation play no part.
This result is of such central significance that attempts are regularly made to understand it afresh and to extend it in other situations involving turbulence.
If we consider the 2D turbulence\cite{Kraichnan}, then in the inviscid limit, we have two conserved quantities -- energy and enstrophy.
This gives rise to two fluxes with the enstrophy flux occurring from the larger to the smaller spatial scales.
The energy flux goes in the reverse direction.
QG turbulence\cite{Charney} stands somewhere in between 2D and 3D turbulences.
In the inviscid limit, besides total energy, QG flows enjoy the possession of yet another conserved quantity which is conserved at the horizontal projection of the particle motion.
We shall call this pseudo-potential vorticity to distinguish it from the potential vorticity that is conserved at a particle in a homentropic fluid.
Defining pseudo-potential enstrophy as half the square of the pseudo-potential vorticity, one would say that like 2D turbulence there are two cascades --- forward cascade of pseudo-potential vorticity and inverse cascade of energy --- in QG turbulence which, however, is inherently three dimensional in nature.
\\
Recently, a paper\cite{Lindborg} has calculated some structure functions in QG turbulence and has made illuminating revelation that isotropy in the sense of Charney\cite{Charney} is useless in deriving the structure functions for QG turbulence.
It has gone on to show that formulation of QG turbulence under the constraint of axisymmetry is productive.
However, it criticized (though somewhat rightly) the ineffectiveness of use of tensorial quantities in the case of QG turbulence in deriving the results.
Now, manipulating the tensorial quantities are at the heart of the derivation of many important two-point velocity correlation functions and other ones\cite{Landau}.
The technique is very intuitive and straightforward.
It has, recently, also been thoroughly used to find out various correlation functions for 2D turbulence\cite{Sagar}.
In this paper, we shall closely (and trickily) follow the original Kolmogorov method put forward in details in the fluid dynamics text due to Landau and Lifshitz\cite{Landau}; and repeated in the ref.-(\cite{Sagar}), to derive structure functions in QG turbulence.
The method has the extra advantage to being able to probe into the form for the two-point third order velocity correlation function in the forward pseudo-potential enstrophy cascade regime --- this has remained untouched in ref.-(\cite{Lindborg}).
\section{Two point third order mixed correlation function}
First of all we shall briefly introduce the necessary equations (see ref.-(\cite{Salmon}) for details).
Let $\vec{u}$ be the three dimensional velocity of the fluid in a frame rotating with constant angular velocity $\vec{\Omega}$.
The fluid body (such as ocean) is assumed to be of uniform density with free surface at $z=\xi(x,y,t)$.
Suppose the bottom $z=-H(x,y)$ is rigid.
The shallow-water equations, then, are:

\be
\f{\pa h}{\pa t}+\vec{\nabla}(\vec{v}h)=0
\label{swe1}\\
\f{D\vec{v}}{Dt}+\vec{f}\times\vec{v}=-g\vec{\nabla}\xi
\label{swe}
\en
Here, $h(x,y,t)\equiv\xi(x,y,t)+H(x,y)$, $\vec{v}\equiv(u_x,u_y)$, $\vec{\nabla}\equiv(\pa_x,\pa_y)$, $\f{D}{Dt}\equiv\f{\pa}{\pa t}+\vec{v}.\vec{\nabla}$ and $\vec{v}=\vec{v}(x,y,t)$.
$f$ is Coriolis parameter that is Taylor-expanded to write $f=f_0+\beta y$.
Using the equations (\ref{swe1}) and (\ref{swe}), one gets the relation:
\be
\f{D}{Dt}\left[\f{\hat{z}.(\textrm{{\bf curl}}\vec{u})+f}{h}\right]=0
\label{swe2}
\en
Let us assume: a) Rossby number $Ro\ll1$, b) Fractional changes in $h$ are small and c) $\beta L/f_0\ll1$ where $L$ is the horizontal scale of the flow.
Imposing these three assumptions on the shallow-water equations one can modify the relation (\ref{swe2}) to yield
\be
\f{\pa q}{\pa t}+\vec{v}.\vec{\nabla}q=0
\label{qqqqq}
\en
where $q=\nabla^2\psi+f-f_0^2\psi/gH_0+f_0(H_0-H)/H_0$ ($\psi$ being $g\xi/f_0$) may be called pseudo-potential vorticity.
Under the same assumptions, for QG flow, one also has the condition:
\be
\vec{\nabla}.\vec{v}=0
\label{qqqq}
\en
Now, the trick is to select an arbitrary two-dimensional plane in the QG turbulent flow such that the plane's normal is parallel to the vertical ({\it i.e.}, along $\vec{f}$) and impose the property of homogeneity and isotropy in the plane only.
By the way, one must keep in mind that the so-called fundamental scale of 3D turbulence has its analogy as the horizontal length scale $L$ for the case of QG turbulence and; the correlation functions to be derived for the forward cascade in this paper are valid in the range (which we shall call inertial range) that is much smaller than $L$ but quite larger than the scale at which the dissipation is effective and, the structure function to be derived for the inverse cascade is valid in the range whose scale is larger than the scale at which energy is being fed in. 
As we shall consider fluid bodies of uniform density only, we shall take density to be unity and let $\vec{\r}$ take over the task of representing position vector in the 2D plane.
Yet another convention: The Greek subscripts used herein can take two values $\r$ and $\p$ which respectively mean the component along the radial vector $\r$ and the component in the transverse direction.
When we shall use the Latin subscript ({\it e.g.}, $a$), it should mean that it can take one more value apart from the ones mentioned above: the third value `$z$' would signify the vertical direction.
Einstein's summation convention will be used extensively.
Also,
\be
\vec{\r}=\vec{\r}_2-\vec{\r}_1,\phantom{xxx}\r^o_\a\equiv\r_\a/{|\vec{\r}|},\phantom{xxx}\r^o_\r=1,\phantom{xxx}\r^o_\p=0
\label{extra}
\en
Now, if $\vec{v}_1$ and $\vec{v}_2$ represent the horizontal fluid velocities at the two neighbouring points at $\r_1$ and $\r_2$ respectively then with similar meaning for $q_1$ and $q_2$, one may define:
\be
K&\equiv&\la\omega_1\omega_2\ra
\label{25}\\
\textrm{and,}\phantom{xxx}\mega&\equiv&\la(\omega_2-\omega_1)(\omega_2-\omega_1)\ra
\label{26}
\en
The angular brackets denote an averaging procedure which averages over all possible positions of points $1$ and $2$ at a given instant of time and a given separation.
Due to homogeneity, $\mega$ may be re-expressed as:
\be
\mega=2\la\omega^2\ra-2K
\label{27}
\en
For simplicity, we shall take a rather idealised situation of QG turbulence which is homogeneous and isotropic on every scale in the plane.
For the unforced case, the component of the correlation tensor will obviously be dependent on time, a fact which won't be shown explicitly in what follows.
As the features of local QG turbulence should be independent of averaged flow, the result derived below is applicable also to the local turbulence in the plane at scale $\r$ much smaller than the fundamental scale.\\
Again, we define a two-point third order mixed correlation tensor in inertial range:
\be
&{}&\mega_\b\equiv\la(v_{2\b}-v_{1\b})(\omega_2-\omega_1)(\omega_2-\omega_1)\ra
\label{32}\\
\R&{}&\mega_\b=4K_{\b}+2M_{\b}
\label{33}
\en
where,
\be
K_{\b}&\equiv&\la v_{1\b}\omega_1\omega_2\ra
\label{34}\\
\textrm{and,}\phantom{xxx}M_{\b}&\equiv&\la\omega_1\omega_1v_{2\b}\ra
\label{35}
\en
Let due to isotropy and homogeneity, we can write following form for $M_{\b}$:
\be
&{}&M_{\b}=M(\r)\r_{\b}^o
\label{36}\\
\R&{}&\f{\pa}{\r_{2\b}}M_{\b}=\la\omega_1\omega_1\pa_{2\b}v_{2\b}\ra=0
\label{37}\\
\R&{}&\f{\pa}{\pa\r}M({\r})+\f{M({\r})}{\r}=0\\
\R&{}& M({\r})=\f{\textrm{constant}}{\r}=0
\label{38}
\en
In the relation (\ref{37}), we are using the expression (\ref{qqqq}) and in writing the relation (\ref{38}) we have taken into account the fact that $M_{\b}$ should remain finite when $\r=0$.
Relations (\ref{36}) and (\ref{38}) imply that:
\be
M_{\b}=0
\label{39}
\en
using which in the relation (\ref{33}), we get:
\be
\mega_{\b}=4K_{\b}
\label{40}
\en
From the equation (\ref{qqqqq}), we may write for the points 1 and 2 respectively:
\be
\f{\pa}{\pa t}\omega_{1}=-v_{1\gamma}\pa_{1\gamma}\omega_{1}
\label{41}\\
\f{\pa}{\pa t}\omega_{2}=-v_{2\gamma}\pa_{2\gamma}\omega_{2}
\label{42}
\en
Multiplying equations (\ref{41}) and (\ref{42}) by $\omega_2$ and $\omega_1$ respectively and averaging subsequently after adding, we get the following differential equation:
\be
\f{\pa}{\pa t}K=2\pa_{\b}K_{\b}
\label{43}
\en
where we have used the fact $\pa_{\b}=-\pa_{1\b}=\pa_{2\b}$.
Using relations (\ref{27}) and (\ref{40}) in the equation ({\ref{43}}), one gets for the inertial range for the pseudo-potential enstrophy cascade in homogeneous and isotropic QG turbulence (forced at an intermediate scale or unforced) in inviscid limit the following differential equation:
\be
&{}&\f{\pa}{\pa t}\la\omega^2\ra-\f{1}{2}\f{\pa}{\pa t}\mega=\f{1}{2\r}\f{\pa}{\pa\r}\left(\r\mega_{\r}\right)
\label{44}\\
\R&{}&\mega_{\r}=-2\eta\r
\label{45}
\en
In getting relation ({\ref{45}}) from the equation (\ref{44}), we have assumed the facts:
\begin{enumerate}
\item $\f{1}{2}\f{\pa}{\pa t}\la\omega^2\ra=-\eta$, {\it i.e.,} there exists a pseudo-potential enstrophy sink at small scales due to some dissipative force such as viscosity and $\eta$ is the finite and constant dissipation rate of the mean pseudo-potential enstrophy.
\item $\f{1}{2}\f{\pa}{\pa t}\mega\approx 0$ due to quasi-stationarity. It may be supposed that the value of $\mega$ varies considerably with time only over an interval corresponding to the fundamental scale of turbulence and in relation to local turbulence the unperturbed flow may be regarded as steady which mean that for local turbulence one can afford to neglect $\f{\pa}{\pa t}\mega$ in comparison with the pseudo-potential enstrophy dissipation rate $\eta$.
\end{enumerate}
\section{Two point third order velocity correlation function}
Again, one may define a rank two correlation tensor:
\be
B_{\a\b}\equiv\la(v_{2\a}-v_{1\a})(v_{2\b}-v_{1\b})\ra
\label{1}
\en
Isotropy and homogeneity in the plane suggests following general form for $B_{\a\b}$
\be
B_{\a\b}=A_1(\r)\delta_{\a\b}+A_2(\r)\r^o_\a\r^o_\b
\label{2}
\en
where $A_1$ and $A_2$ are functions of time and $\r$.
Making use of the relations (\ref{extra}) in the equation (\ref{2}), one gets:
\be
B_{\a\b}=B_{\p\p}(\delta_{\a\b}-\r^o_\a\r^o_\b)+B_{\r\r}\r^o_\a\r^o_\b
\label{3}
\en
One may break the relation (\ref{1}) as
\be
B_{\a\b}=\la v_{1\a}v_{1\b}\ra+\la v_{2\a}v_{2\b}\ra- \la v_{1\a}v_{2\b}\ra-\la v_{2\a}v_{1\b}\ra
\label{4}
\en
Defining
\be
b_{\a\b}\equiv\la v_{1\a}v_{2\b}\ra
\label{5}
\en
one may proceed, keeping in mind the isotropy and the homogeneity, to write
\be
B_{\a\b}=\la v^2\ra\delta_{\a\b}-2b_{\a\b}
\label{6}
\en
Again, having used the condition (\ref{qqqq}), one may write:
\be
&{}&\pa_\b B_{\a\b}=0\nonumber\\
\Rightarrow&{}& B'_{\r\r}+\f{1}{\r}(B_{\r\r}-B_{\p\p})=0\nonumber\\
\Rightarrow&{}&B_{\p\p}=\r B'_{\r\r}+B_{\r\r}
\label{7}
\en
where the equation (\ref{3}) has been used and prime ($'$) denotes derivative w.r.t. $\r$.
Let's give yet another definition:
\ben
b_{\a\b,\gamma}\equiv\la v_{1\a}v_{1\b}v_{2\gamma}\ra
\enn
Invoking homogeneity and isotropy in the plane once again along with the symmetry in the first pair of indices, one may write the most general form of the third rank Cartesian tensor for this case as
\be
b_{\a\b,\gamma}&=&C(\r)\delta_{\a\b}\r^o_\gamma+D(\r)(\delta_{\gamma\b}\r^o_\a+\delta_{\a\gamma}\r^o_\b)+F(\r)\r^o_\a\r^o_\b\r^o_\gamma
\label{10}
\en
where, $C$, $D$ and $F$ are functions of $\r$.
Yet again, the expression (\ref{qqqq}) dictates:
\be
\f{\pa}{\pa_{2\gamma}}b_{\a\b,\gamma}=\f{\pa}{\pa_{\gamma}}b_{\a\b,\gamma}=0\nonumber\\
\Rightarrow C'\delta_{\a\b}+\f{C}{\r}\delta_{\a\b}+\f{2D}{\r}\delta_{\a\b}+\f{2D'}{\r^2}\r_\a\r_\b-\f{2D}{\r^3}\r_\a\r_\b+\f{F'}{\r^2}\r_\a\r_\b+\f{F}{\r^3}\r_\a\r_\b=0
\label{11}
\en
Putting $\a=\b$ in equation (\ref{11}) one gets:
\be
2C+2D+F=\f{\textrm{constant}}{\r}=0
\label{11-12}
\en
where, it as been imposed that $b_{\a\b,\gamma}$ should remain finite for $\r=0$.
Again, using equation (\ref{11}), putting $\a\ne\b$ and manipulating a bit one gets:
\be
D=-\f{1}{2}(\r C'+C)
\label{12}
\en
using which in relation (\ref{11-12}), one arrives at the following expression for $F$:
\be
F=\r C'-C
\label{13}
\en
Defining
\be
B_{\a\b\gamma}&\equiv&\la(v_{2\a}-v_{1\a})(v_{2\b}-v_{1\b})(v_{2\gamma}-v_{1\gamma})\ra\nonumber\\
&=&2(b_{\a\b,\gamma}+b_{\gamma\b,\a}+b_{\a\gamma,\b})
\label{13-14}
\en
and putting relations (\ref{12}) and (\ref{13}) in the equation (\ref{13-14}) and using relation (\ref{10}), one gets:
\be
&{}&B_{\a\b\gamma}=-2\r C'(\delta_{\a\b}\r^o_\gamma+\delta_{\gamma\b}\r^o_\a+\delta_{\a\gamma}\r^o_\b)+6(\r C'-C)\r^o_\a\r^o_\b\r^o_\gamma\\
\label{14}
\Rightarrow&{}& B_{\r\r\r}=-6C
\label{15}
\en
which along with relations (\ref{12}), (\ref{13}) and (\ref{10}) yields the following expression:
\be
&{}&b_{\a\b,\gamma}=-\f{B_{\r\r\r}}{6}\delta_{\a\b}\r^o_\gamma+\f{1}{12}(\r B'_{\r\r\r}+B_{\r\r\r})(\delta_{\gamma\b}\r^o_\a+\delta_{\a\gamma}\r^o_\b)-\f{1}{6}(\r B'_{\r\r\r}-B_{\r\r\r})\r^o_\a\r^o_\b\r^o_\gamma
\label{16}
\en
The equation (\ref{swe}) suggests:
\be
\f{\pa}{\pa t}v_{1\a}=-v_{1\gamma}\pa_{1\gamma}v_{1\a}+f_{1a}\epsilon_{a\a\gamma}v_{1\gamma}-g\pa_{1\a}\xi_1
\label{17}\\
\f{\pa}{\pa t}v_{2\b}=-v_{2\gamma}\pa_{2\gamma}v_{2\b}+f_{1a}\epsilon_{a\b\gamma}v_{2\gamma}-g\pa_{2\b}\xi_2
\label{18}
\en
multiplying equations (\ref{17}) and (\ref{18}) with $v_{2\b}$ and $v_{1\a}$ respectively and adding subsequently, one gets the following:
\be
\f{\pa}{\pa t}\la v_{1\a}v_{2\b}\ra&=&-\pa_{1\gamma}\la v_{1\gamma}v_{1\a}v_{2\b}\ra-\pa_{2\gamma}\la v_{2\gamma}v_{1\a}v_{2\b}\ra\nonumber\\
&{}&+\epsilon_{a\a\gamma}\la f_{1a}v_{1\gamma}v_{2\b}\ra+\epsilon_{a\b\gamma}\la f_{2a}v_{2\gamma}v_{1\a}\ra\nonumber\\
&{}&-g\pa_{1\a}\la \xi_1 v_{2\b}\ra-g\pa_{2\b}\la \xi_2 v_{1\a}\ra
\label{19}
\en
Due to isotropy, the correlation function $\la \xi_1\vec{v}_2\ra$ should have the form $f(\r)\vec{\r}/|\vec{\r}|$.
But since, $\pa_{\a}\la \xi_1{v}_{2\a}\ra=0$ due to solenoidal velocity field $f(\r)\vec{\r}/|\vec{\r}|$ must have the form $\textrm{constant}\times(\vec{\r}/|\vec{\r}|^2)$ that in turn must vanish to keep correlation functions finite even at $\r=0$.
Thus, equation (\ref{19}) can be written as:
\be
\f{\pa}{\pa t}b_{\a\b}=\pa_{\gamma}(b_{\a\gamma,\b}+b_{\b\gamma,\a})+f_0\epsilon_{z\a\gamma}b_{\gamma\b}+f_0\epsilon_{z\b\gamma}b_{\a\gamma}
\label{20}
\en
Here we have used the approximation: $\vec{f}=f_0\hat{z}$.
Using equations (\ref{6}) and (\ref{16}), one can rewrite equation (\ref{20}) as:
\be
\f{1}{2}\f{\pa}{\pa t}\la v^2\ra-\f{1}{2}\f{\pa}{\pa t}B_{\r\r}=\f{1}{6\r^3}\f{\pa}{\pa \r}\left(\r^3B_{\r\r\r}\right)
\label{21}
\en
Note that the terms containing the Levi-Civita symbol vanish of the joint effect of the expressions (\ref{3}) and (\ref{6}); and the antisymmetry property of Levi-Civita symbol.
As we are interested in the pseudo-potential enstrophy cascade, the first term in the L.H.S. is zero because of energy remains conserved in QG turbulence in the inviscid limit; it cannot be dissipated at smaller scales.
Also, as we are interested in the forward cascade which is dominated by pseudo-potential enstrophy cascade, on the dimensional grounds in the inertial range $B_{\r\r}$ (if it is assumed to depend only on $\eta$ and $\r$) may be written as:
\be
\f{\pa}{\pa t}B_{\r\r}=\Gamma\eta\r^2
\label{22}
\en
where $\Gamma$ is a numerical proportionality constant.
Hence, using the relation (\ref{22}), the equation (\ref{21}) reduces to the following differential equation:
\be
\f{1}{6\r^3}\f{\pa}{\pa \r}\left(\r^3B_{\r\r\r}\right)=-\f{\Gamma}{2}\eta\r^2
\label{23}
\en
which when solved using relation(\ref{15}) imposing finiteness of $B_{\r\r\r}$ for $\r=0$, one gets
\be
B_{\r\r\r}=-\f{\Gamma\eta}{2}\r^3
\label{24}
\en
The relation (\ref{24}) is the expression for the two-point third order correlation function in the isotropic and homogeneous plane of QG turbulence (forced or unforced) in the range of the forward cascade where there is no overlapping with energy cascade.
Since $\Gamma$ has not been determined one must confess that the equation (\ref{24}) is just a scaling law at this stage.
\\
Now suppose the fluid body is being forced at small scales {\it i.e.}, energy is being supplied and the mean rate of injection of energy per unit mass is denoted by $\varepsilon_u$ (assumed finite and constant).
Let us focus on the inverse energy cascade.
Then technically we have to proceed just as before to finally arrive at the differential equation (\ref{21}).
One obviously would set $\f{1}{2}\f{\pa}{\pa t}\la v^2\ra=\f{2}{3}\varepsilon_u$ invoking the hypothesis\cite{Charney} that there should be equipartition of energy between potential energy and the energy content in each of the two horizontal velocity components in the plane.
Lets also assume that $\f{\pa}{\pa t}B_{\r\r}\approx 0$ in the inverse cascade regime supposing the forced QG turbulence to be in the state of quasi-stationarity.
So we are left with the following differential equation:
\be
&{}&\f{1}{6\r^3}\f{\pa}{\pa \r}\left(\r^3B_{\r\r\r}\right)=\f{2}{3}\varepsilon_u
\label{derte}\\
\R&{}&B_{\r\r\r}=+\varepsilon_u\r
\label{energy_cascade_S3}
\en
where in the last step the integration constant has been set to zero to prevent $B_{\r\r\r}$ from blowing up at $\r=0$.
The expression (\ref{energy_cascade_S3}) is the expression for the two-point third order correlation function in the isotropic and homogeneous plane of forced QG turbulence for the inverse energy cascade. 
\section{Conclusions and Discussions}
It has, thus, been again showcased how handy and useful the Kolmogorov-Landau approach can prove to be.
The results (\ref{45}) and (\ref{energy_cascade_S3}) naturally agree with what has been arrived at by Lindborg\cite{Lindborg} earlier.
Within the domain of the approximations made these results are exact, something worth getting as the literature of turbulence is comparatively barren as far as exact relations are concerned.
However, the hypothesis of the equipartition of energy used in equation (\ref{derte}) is as questionable as the assumption of isotropy in the sense of Charney.
This hypothesis needs to be put on more firm basis.
In the closing, we hope that validity of the results derived will be checked both numerically and experimentally in near future to see if the approximations made for the homogeneous QG turbulence in this paper are correct or not.
Also, the fact that the structure functions for the inherently three-dimensional QG turbulence are more like that of the 2D turbulence than that of the 3D turbulence speaks volumes for the importance of study of third order structure functions for demystifying the two-dimensionalisation effect\cite{SagarE1, SagarE2} of the 3D turbulent fluid due to rapid rotation. 
\\
\\
The author would like to acknowledge his supervisor --- Prof. J. K. Bhattacharjee --- for the helpful discussions. Also, CSIR (India) is gratefully acknowledged for awarding fellowship to the author. Dr. E. Lindborg has encouraged the author to take up this work by sending his own paper\cite{Lindborg}; he is heartily thanked.
%%%%%%%%%%%%%%%%%%%%%%%%%%%%%%%%%%%%%%%%%%%%%%%%%%%%%%%%%%%%%%%%%%%%%%%%%%%%%%%%%%%%%%%%%%%%%%%%%%%%%%%%%%%%%%%%%%%%%%%%%%%%

%%%%%%%%%%%%%%%%%%%%%%%%%%%%%%%%%%%%%%%%%%%%%%%%%%%%%%%%%%%%%%%%%%%%%%%%%%%%%%%%%%%%%%%%%%%%%%%%%%%%%%%%%%%%%%%%%%%%%%%%%%%%


\begin{thebibliography}{99}
%
\bibitem{Kolmogorov} A. N. Kolmogorov, {\it Dissipation of the energy in the locally isotropic turbulence}, {Dokl. Akad. Nauk SSSR, {\bf{32}}}, 1(1941); (English translation: {Proc. R. Soc. Lond. A {\bf{434}}}, 15 (1991)).
%
\bibitem{Kraichnan} R. H. Kraichnan, {\it Inertial-range transfer in two-and three-dimensional turbulence},{ J. Fluid Mech. {\bf {47}}}, { J. Fluid Mech. {\bf {47}}}, 525 (1971).
%
\bibitem{Charney} J.G. Charney, {\it Geostrophic turbulence}, {J. Atmos. Sci. {\bf 28}}, 1087 (1971)
%
\bibitem{Lindborg} E. Lindborg, {\it Third-order structure function relations for quasi-geostrophic turbulence}, { J. Fluid Mech. {\bf {572}}}, 255 (2007).
%
\bibitem{Landau}  L. D. Landau and E. M. Lifshitz, {\it Fluid Mechanics, Second Edition: Volume 6 (Course of Theoretical Physics)}, (Reed Educational and Professional Publishing Ltd), (1987).
%
\bibitem{Sagar} S. Chakraborty, {\it On the use of Kolmogorov-Landau approach in deriving various correlation functions in 2-D incompressible turbulence}, {Phys. of Fluids \bf{19}}, 085110 (2007)
%
\bibitem{Salmon}R. Salmon, {\it Lectures on Geophysical Fluid Dynamics}, (Oxford University Press, New York), (1998).
%
\bibitem{SagarE1} S. Chakraborty, {\it Signatures of two-dimensionalisation of 3D turbulence in the presence of rotation}, {Europhys. Lett. {\bf 79}}, 14002 (2007).
%
\bibitem{SagarE2} S. Chakraborty and J.K. Bhattacharjee, {\it Third-order structure function for rotating three-dimensional homogeneous turbulent flow}, {Phys. Rev. E {\bf 76}}, 036304 (2007).
\end{thebibliography}
\end{document}